\documentclass[twoside,fleqn]{ActaStyle}
\usepackage{times}

\usepackage{lscape}
\usepackage{graphicx,epsfig}
\usepackage{amsmath,amssymb}

\newcommand{\fm}{~\mathrm{fm}}
\newcommand{\vlk}{$V_{\text{low k}}$}
\newcommand{\vnn}{$V_{\text{NN}}$}
\newcommand{\la}{\Lambda}
\newcommand{\q}{{\bf q}}
\newcommand{\qp}{{\bf q'}}
\newcommand{\pp}{{\bf p}+{\bf p'}}
\newcommand{\ip}{{\bf p''}}
\newcommand{\spin}{{\bf \sigma} \cdot {\bf \sigma'}}

\def\authorheading{A. Schwenk et al.}
\def\shorttitle{RG Approach to Nuclear Physics}

\begin{document}

\pagerange{1}{8}

\title{THE NUCLEON INTERACTION AND NEUTRON MATTER FROM THE
RENORMALIZATION GROUP}

\author{A.~Schwenk\,\footnote{Invited talk
given by A.S. at 5th Int. Conf. Renormalization Group 2002,
Tatransk\'a  \v{S}trba (Slovakia), March, 2002\\
E-mail address: aschwenk@nuclear.physics.sunysb.edu}$^*$,
B.~Friman\,$^\dagger$, S.K.~Bogner\,$^*$,
G.E.~Brown\,$^*$, and T.T.S.~Kuo\,$^*$}
{$^*$Department of Physics and Astronomy, State
University of New York,\\
Stony Brook, N.Y. 11794-3800, U.S.A.\\
$^\dagger$Gesellschaft f\"ur Schwerionenforschung, Planckstr. 1, 64291
Darmstadt, Germany}

\day{June 3, 2002}

\abstract{We show that the renormalization group decimation of
modern nucleon potential models to low momenta results in a unique
nucleon interaction \vlk. This interaction is free of short-ranged
singularities and can be used directly in many-body calculations.
The RG scaling properties follow directly from the invariance of
the scattering phase shifts. We discuss the RG treatment
of Fermi liquids. The RG equation for the scattering amplitude in 
the two particle-hole channels is given at zero temperature.
The flow equations are simplified by retaining only the leading 
term in an expansion in small momentum transfers. The RG flow
is illustrated by first studying a system of spin-polarized 
fermions in a simple model. Finally, results for neutron matter
are presented by employing the unique low momentum interaction \vlk~as 
initial condition of the flow. The RG approach yields the amplitude
for non-forward scattering, which is of great interest for 
calculations of transport properties and superfluid gaps in neutron 
star interiors. The methods used can also be applied to condensed matter 
systems in the absence of long-ranged interactions.}


\section{Introduction}
\setcounter{section}{1} \setcounter{equation}{0}

Over the past decade, there has been a great effort in deriving
low-energy effective nucleon-nucleon interactions~\cite{Encyc}.
Concurrently in the condensed matter community, Shankar proposed
renormalization group methods to the study of strongly interacting
Fermi systems~\cite{Shankar}. In this talk, we discuss how the RG
can be used for the two-nucleon system as well as the nuclear
many-body problem leading to quantitatively quite encouraging
results. The two problems are similar from the RG point of view. 
The degrees of freedom are separated by a cutoff into low and high 
momenta. As the cutoff is decreased, the effects of the high 
momentum modes on the low momentum observables are integrated out 
into effective couplings. For the two-body problem, this 
corresponds to solving the Lippmann-Schwinger equation in the 
particle-particle channel, whereas for the many-body system, it 
constitutes a genuine two- or three-channel problem. Taking the
cutoff to zero, one recovers the scattering length in the two-body 
case and the quasiparticle interaction of Fermi liquid theory in
the matter case.

The RG approach allows us to evolve the full in-medium scattering 
amplitude on the Fermi surface from the vacuum two-body 
interaction. Compared to traditional microscopic calculations for 
Fermi systems, it is relatively straightforward to retain all
momentum scales in the RG. In addition, our method 
is physically more transparent, since the effects of the high 
momentum modes, i.e., scattering to states far away from the
Fermi surface, are completely tractable. 

\section{RG evolution of nucleon interaction models} 

\begin{figure}[t]
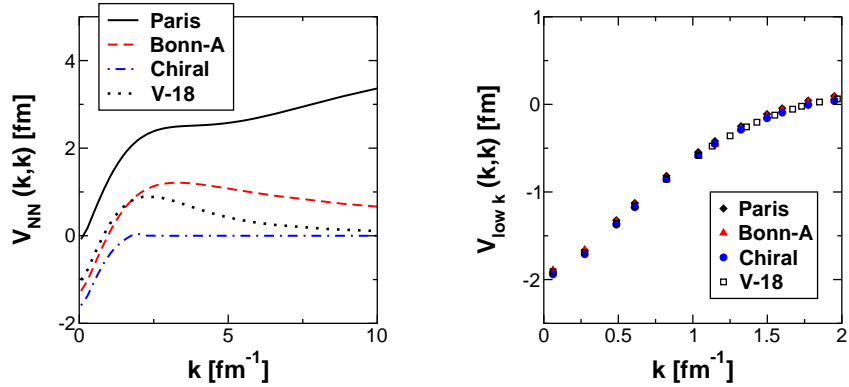

\begin{center}
\includegraphics[scale=0.27,clip=]{4barediag1s0.eps}
\hspace*{1.0cm}
\includegraphics[scale=0.27,clip=]{4vlkdiag2.eps}
\end{center}
\caption{Diagonal matrix elements of several realistic
\vnn~and \vlk~as a function of
relative momentum in the $^1$S$_0$ partial wave. The \vlk~results are
given for a cutoff $\la = 2.0\fm^{-1}$. A comparison of phase shifts
calculated from \vnn~and \vlk~is given in~\cite{Vlowkflow}.}
\label{vlowk}
\end{figure}

We start from a realistic ``bare'' nucleon-nucleon potential \vnn,
which is based e.g., on multiple Yukawa interactions, where the
couplings are fitted to the scattering phase shifts. By experiment,
the various \vnn~are constrained only up to a relative momentum
scale of $k \sim 2.0\fm^{-1}$ (corresponding to a lab energy
$E_{\text{Lab}} \sim 330~\mathrm{MeV}$) and consequently have quite
different momentum components as we show in Fig.~\ref{vlowk}. We
integrate out the unrestricted high momentum modes with the
requirement that the effective potential reproduces the phase shift
data as well as the long range wave function tails, i.e., the
so-called half-on-shell (HOS) $T$
matrix~\cite{Vlowk,Vlowkflow}.\footnote{The invariance of the HOS
$T$ matrix was proposed by Bogner, Kuo and Coraggio~\cite{Vlowk}.
Strictly speaking, only the fully on-shell amplitude is observable.
The invariance of the phase shifts $T(k,k;k^2) = -
\tan\delta / k$ (without any constraints on the HOS $T$ matrix)
can be achieved by a similarity transformation on \vlk. The resulting
low momentum interaction remains unique and the diagonal matrix
elements are basically unchanged. The RG equation is then replaced 
by a symmetrized version of the current one, Eq.~(\ref{rgevlowk}).} 
The HOS scattering amplitude is given by
\begin{equation}
\label{fulltmat}
T(k',k;k^{2}) = V_{\text{NN}}(k',k) + \frac{2}{\pi} \, \mathcal{P}
\int_{0}^{\infty} \frac{V_{\text{NN}}(k',p) \,
T(p,k;k^{2})}{k^{2}-p^{2}} \, p^{2} dp .
\end{equation}

After a resummation of the fast modes and subsequently removing the
so-generated energy-dependence (for details see~\cite{VlowkRG}), the
bare potential can be replaced by the cutoff-dependent \vlk
with slow mode intermediate states only
\begin{equation}
T(k',k;k^{2})= V_{\text{low k}}(k',k) + \frac{2}{\pi} \, \mathcal{P}
\int_{0}^{\Lambda} \frac{V_{\text{low k}}(k',p) \,
T(p,k;k^{2})}{k^{2}-p^{2}} \, p^{2} dp .
\label{lowtmat}
\end{equation}
Alternatively, Eq.~(\ref{lowtmat}) may be used as a definition of \vlk.

The scaling properties of \vlk~follow directly from the invariance $d
T(k',k;k^{2}) /d \Lambda =0$. This leads to the RG equation for
energy-independent effective interactions~\cite{VlowkRG}
\begin{equation}
\frac{d}{d \Lambda} V_{\text{low k}}(k',k) = \frac{2}{\pi}
\frac{V_{\text{low k}}(k',\Lambda) \: T(\Lambda,k;\Lambda
^{2})}{1-(k / \Lambda)^{2}} .
\label{rgevlowk}
\end{equation}
A similar RG equation for energy-dependent quasi-potentials
$Q(k',k;\omega)$ was derived by Birse {\em et al.}~\cite{Birse}.
In both cases, there exists a non-trivial fixed point
corresponding to an infinite scattering length or a bound state at
threshold~\cite{Birse,Perry}.

\begin{figure}[t]
\begin{center}
\includegraphics[scale=0.27,clip=]{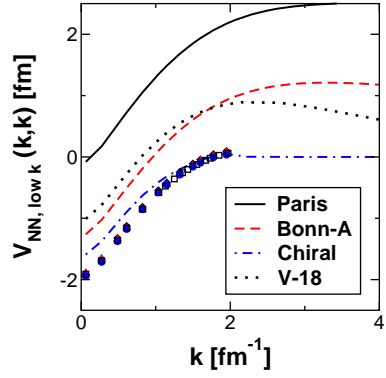}
\end{center}
\caption{The bare potential models compared to the unique \vlk. As in
Fig.~\ref{vlowk}, we show the diagonal matrix elements in partial wave
$^1$S$_0$.}
\label{shift}
\end{figure}
Starting at large cutoffs, the RG flow is used to evolve the bare
potentials to low momenta. We find that, for cutoffs below the
scale set by the experimental data, $\la \sim 2.0\fm^{-1}$, the RG
decimation leads to an unambiguous low momentum potential
\vlk~\cite{Vlowk,Vlowkflow}, independent of the initial bare potential, 
as is shown in Fig.~\ref{vlowk}. Similar
results are found in all partial waves. The resulting interaction
can be used as a benchmark for the interaction derived in
chiral effective field theories. By comparing with the bare
interaction, Fig.~\ref{shift}, we observe that the main effect of
the RG evolution is a constant shift in momentum space (see
also~\cite{MB11}).

\begin{figure}[t]
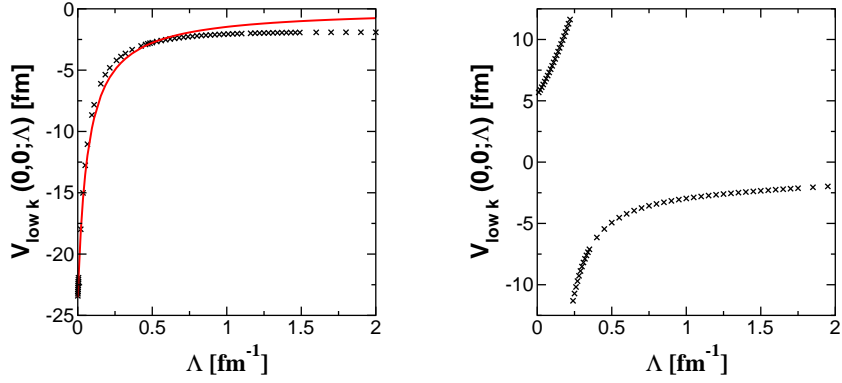

\begin{center}
\includegraphics[scale=0.27,clip=]{1s0lambdarg.eps}
\hspace*{1.0cm}
\includegraphics[scale=0.27,clip=]{3s1lambda.eps}
\end{center}
\caption{The flow of $V_{\text{low k}}(0,0)$ (crosses) with 
the cutoff in partial waves $^1$S$_0$ (left) and $^3$S$_1$ (right figure).}
\label{vlowkrg}
\end{figure}

Therefore, it is instructive to follow the flow of the
interaction at zero relative momentum, $V_{\text{low k}}(0,0)$,
with the cutoff. This is shown in Fig.~\ref{vlowkrg}.
We observe that, in the $^1$S$_0$ channel there is a clear
separation of scales. For cutoffs in the range between the
mass of the lightest exchange meson, $m_\pi \approx 0.7~\fm^{-1}$,
and the mass scale corresponding to the short-ranged repulsion,
$m_\omega \approx 4~\fm^{-1}$, the interaction is basically constant.
On the other hand, in the $^3$S$_1$ channel, where the tensor force 
contributes, a weak dependence on the cutoff reflects a less
precise separation of scales due to higher order tensor 
interactions. As the flow is evolved to $\la = 0$, 
\vlk~reproduces correctly the scattering lengths ($a_{^1 S_0} = - 
23.71\fm$ and $a_{^3 S_1} = 5.425\fm$). The jump in the $^3$S$_1$
channel accounts for the compensation of the deuteron bound state.
For cutoffs $\displaystyle \Lambda < \sqrt{m \, E_{\text{D}}}$, 
where $m$ denotes 
the nucleon mass and $E_{\text{D}}$ is the deuteron binding energy, 
the bound state is integrated out and must be included in the 
effective potential~\cite{Vlowkflow}.

In Fig.~\ref{vlowkrg} we also compare the exact \vlk~result
with a flow prediction for small cutoffs. By retaining only the 
leading term in the effective range expansion, one obtains an 
analytic solution to the RG equation
\begin{equation}
V_{\text{low k}}(0,0) = \frac{1}{1 / a_{^1 S_0} - 
2 \Lambda / \pi}~~~\text{as}~~~\la \to 0 ,
\end{equation}
which is shown as solid line in Fig.~\ref{vlowkrg}. The agreement
between the approximate solution and \vlk~is very good for $k <
1/|a_{^1 S_0}|$.

The constant shift of \vlk~relative to the bare potentials
corresponds (due to the cutoff) to a smeared delta-function and 
removes the short-ranged repulsion from the bare potential. 
Furthermore, many-body effects can be systematically included. 
Consequently, \vlk~is an attractive alternative to the Brueckner 
$G$ matrix in microscopic many-body calculations.

\section{RG for Fermi liquids in the particle-hole channels} 

RG methods can be applied to the nuclear many-body problem using
the approach proposed by Shankar~\cite{Shankar}. We employ the RG
equation in the particle-hole channels to generate the full
scattering amplitude on the Fermi surface from the two-body
interaction. To this end, we impose a cutoff on the particle and
hole modes around the Fermi surface. We integrate out 
the fast particles and fast holes with momenta $|p-k_{\text{F}}| > \la$. 
The RG is initialized with the vacuum two-body
interaction \vlk~at initial cutoff $\la_0 = k_{\text{F}}$. The interaction
between particles on the Fermi surface is not renormalized by
particle-hole polarization effects with momentum transfers larger
than $2 \, \la_0$ (outside the initial cutoff shell). By evolving
the scattering amplitude down to the Fermi surface, i.e., $\la = 
0$, we generate the full scattering amplitude for low energy 
excitations. In particular, we reproduce the Fermi liquid theory 
relations among the quasiparticle interaction and the forward 
scattering amplitude.\footnote{Thus, the important contributions
from low-lying particle-hole excitations to the effective 
interaction are explicitly taken into account. The main effect of 
scattering in the particle-particle channel, the removal of the 
short-ranged repulsion, is taken care of by using \vlk~as the 
starting point for the RG flow. The low-lying excitations in this 
channel, which are responsible e.g., for superfluidity, are not
included. These may be treated explicitly by employing BCS theory 
for the quasiparticle scattering amplitude to compute the superfluid gap.}
\begin{figure}[t]
\begin{center}
\includegraphics[scale=0.7,clip=]{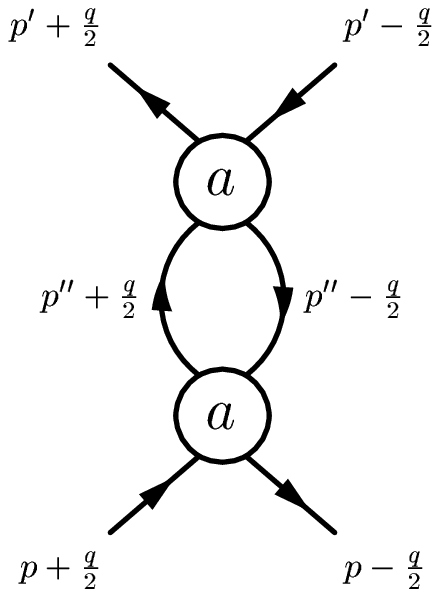}
\hspace*{0.3cm}
\includegraphics[scale=0.7,clip=]{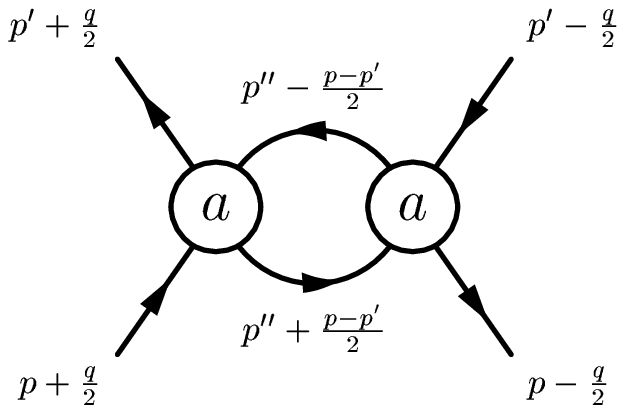}
\hspace*{0.2cm}
\includegraphics[scale=0.23,clip=]{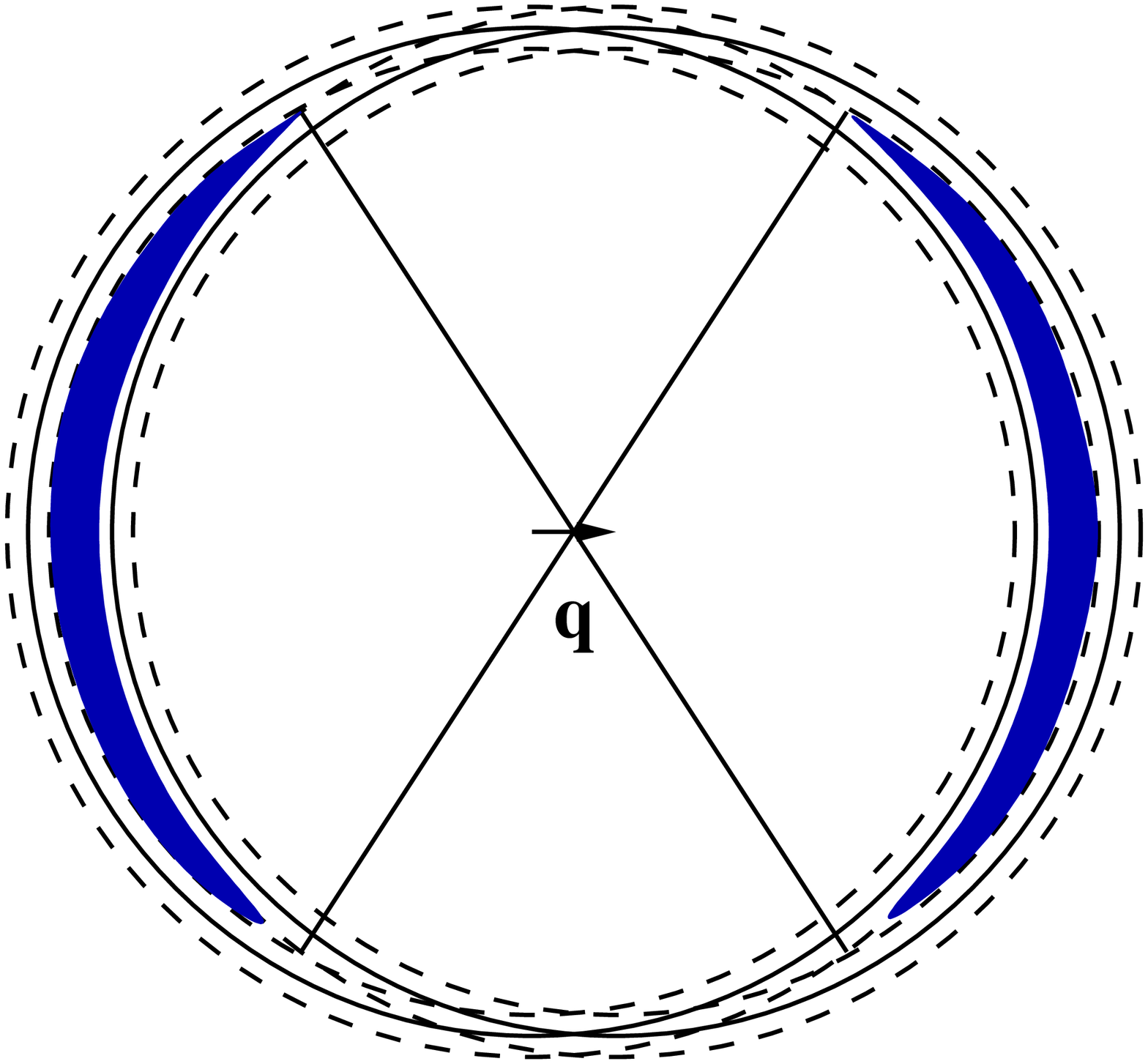}
\end{center}
\caption{The one-loop contributions to the beta functions.
The Pauli principle requires to retain both the zero sound
(ZS) channel with momentum transfer $\q$ in the particle-hole loop
(left figure) and the exchange channel (ZS') with momentum transfer
$\qp = {\bf p} - {\bf p'}$ (middle figure). The right figure shows the fast
particle-hole contributions (filled shells) to the ZS flow. The
dashed circles denote the cutoffs around the Fermi surface for the
particle-hole pair in the loop.}
\label{channels}
\end{figure}

The Fermi liquid RG is at least a two channel problem. This is
mandated by the Pauli principle~\cite{BB,Dupuis}. The RG equations
at one-loop receive contributions from the forward (ZS) and
exchange (ZS') channel, which are depicted in Fig.~\ref{channels}.
At the one-loop level, the running of the amplitude originates from
the available fast particle-hole states only. The RG equation
is antisymmetric by construction and reads~\cite{IIpaper} (the spin
variables are suppressed for simplicity)
\begin{align}
\frac{d}{d \Lambda} a(\q,\qp) & = \frac{d}{d \Lambda} \biggl\{ \:
g \int\limits_{\text{fast}, \Lambda} \frac{d^3 \ip}{(2 \pi)^3} \:
\frac{n_{\ip+\q/2}-n_{\ip-\q/2}}{\varepsilon_{\ip+\q/2}-\varepsilon_{\ip-\q/2}}
\biggr\} \: a\bigl(\q,\frac{\pp}{2}+\frac{\qp}{2}-\ip\bigr) \nonumber \\
& \times a\bigl(\q,\ip-\frac{\pp}{2}+\frac{\qp}{2}\bigr) - \bigl\{ \q
\leftrightarrow \qp \bigr\} ,
\end{align}
where $g$ is the spin-isospin degeneracy and the momenta are defined in
Fig.~\ref{channels}.
A resummation of the beta functions in the particle-hole channels
is performed in~\cite{IIpaper}. In short, it generates $df/d\la$
vertices in the beta function, where the quasiparticle interaction
$f$ receives contributions only from the ZS-irreducible flow (for
details see~\cite{IIpaper}). This corresponds to the running of the
vertex in one channel while integrating out fast particle-hole
contributions in the second channel.

The phase space contributing to the flow, e.g., in the ZS
channel, is depicted in Fig.~\ref{channels}. We employ an
expansion in momentum transfers to solve the RG equation. It is
based on the observation that the dependence on Landau angle in the 
ZS' flow (the ``induced interaction'' \cite{BB}) is small. For 
$q,q' \ll k_{\text{F}}$, the beta functions may be obtained
analytically. This leads to a flow equation, which in a
schematic notation, is of the form
\begin{equation}
\frac{d}{d \Lambda} a(\q,\qp) = \Theta(q-2\la) \:
\beta_{\text{ZS}}[a,\q,\la] - \Theta(q'-2\la) \:
\beta_{\text{ZS'}}[a,\qp,\la] ,
\end{equation}
where $\beta_{\text{ZS}^{(\prime)}}[a,{\bf q^{(\prime)}},\la]$
is derived in detail in~\cite{RGnm}.
We note that the flow starts for $\la \leq q^{(\prime)}/2$ (see
also Fig.~\ref{channels}). This can be used to switch off either
the ZS or ZS' contributions. E.g., by setting $q=0$, we retain
only the exchange contributions. These correspond to the induced
interaction of Babu and Brown~\cite{BB}.\footnote{We note that the
RG approach yields, in a Pauli principle conserving approximation, 
the effective interaction and scattering amplitude for finite 
scattering angles. In fact, to the best of our knowledge, it 
constitutes the only microscopic approach where both momentum 
transfers are treated on an equal footing.} We use this observation 
to define the quasiparticle interaction and the forward scattering 
amplitude of Fermi liquid theory (FLT) in the RG approach,
\begin{equation}
f_{\text{FLT}}(\qp) = \lim_{\la \to 0} a(q=0,\qp,
\la)~~~~~\text{and}~~~~~a_{\text{FLT}}(\qp) 
= \lim_{q \to 0} a(\q,\qp,\la=0) .
\end{equation}
This ambiguity in the limiting procedure corresponds to the long
wavelength-low energy singularity in the particle-hole channel,
which was used by Landau to eliminate the ZS contribution.

\section{Results for a toy model and neutron matter} 

We demonstrate the interplay of the particle-hole channels at the
one-loop level for a toy system of spin-polarized fermions. 
The RG is solved at zero temperature and for three dimensions in the 
small-momentum transfer approximation ($q,q' \ll k_{\text{F}}$). The complete
phase space in two dimensions was studied in~\cite{BF}.
\begin{figure}[h]
\begin{center}
\includegraphics[scale=0.27,clip=]{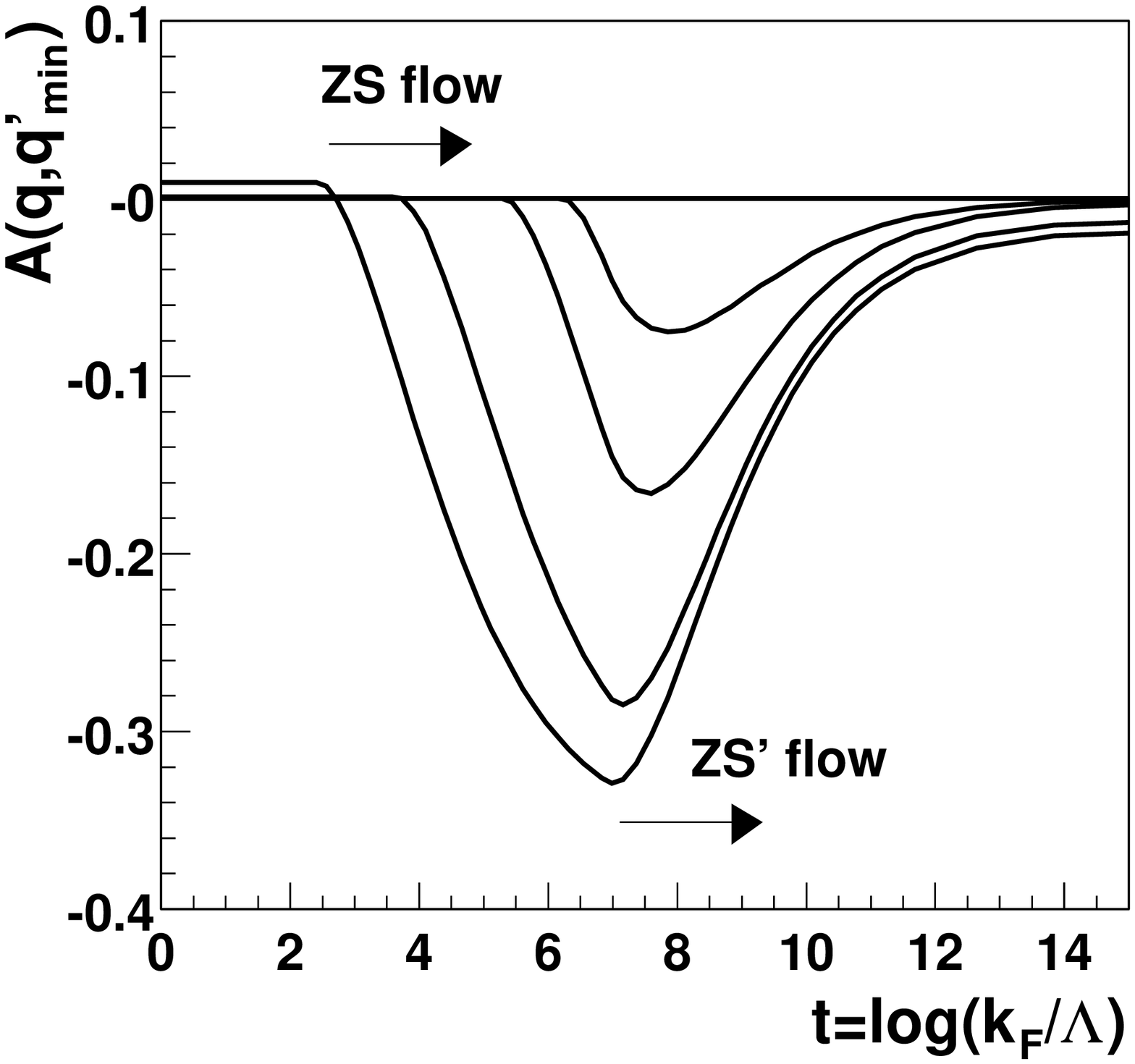}
\hspace*{0.5cm}
\includegraphics[scale=0.27,clip=]{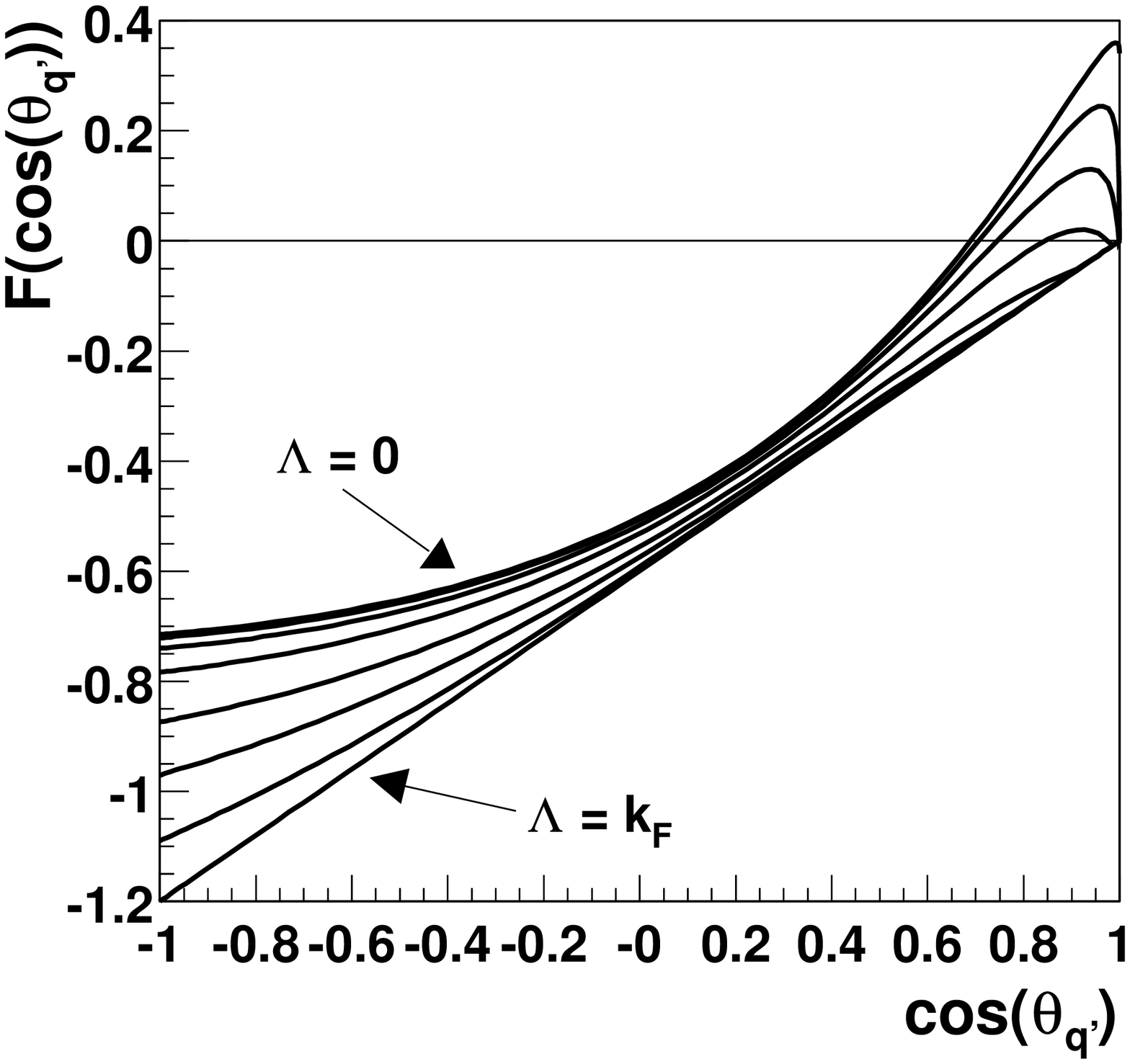}
\end{center}
\caption{Interplay of the particle-hole channels at one-loop. The left
figure tracks the flow of $A(q,q'_{\text{min}}=0.002 k_{\text{F}})$ versus
$t=\log(k_{\text{F}}/\la)$ for several values of $q \geq q'_{\text{min}}$. 
The right figure shows snapshots of the running of the quasiparticle 
interaction $F(\cos\theta_\qp)$ as function of $\cos\theta_\qp$ at
different cutoffs.}
\label{toyflow}
\end{figure}
We choose the initial two-body interaction to be 
(in units where the density of states $m \, k_{\text{F}} / 2 \pi^2 = 1$)
\begin{equation}
a(\q,\qp,\la=k_{\text{F}}) = \frac{1}{2} \bigl( \cos\theta_\qp - \cos\theta_\q
\bigr) ,~~~\text{where}~~~q^{(\prime)} = k_{\text{F}} \, \sqrt{2 - 2
\cos\theta_{\q^{(\prime)}}}
\label{direct}
\end{equation}
and the angle $\theta_\qp$ reduces to the Landau angle for $q = 0$.
This toy interaction has only $F_0 = -6/10$ and $F_1 = 6/10$
non-vanishing Fermi liquid parameters.

In Fig.~\ref{toyflow}, we show the running of the scattering
amplitude and the quasiparticle interaction with $t =
\log(k_{\text{F}}/\la)$. The left graph of Fig.~\ref{toyflow} shows the
interference of the ZS and the ZS' channel. The phase
space is open for $\la \leq q,q'/2$. Accordingly we observe the
onset of the ZS channel first in the running of
$A(q,q'_{\text{min}})=m^\star/m \, a(q,q'_{\text{min}})$. 
The ZS' channel sets in later and counteracts 
the ZS channel, as the two channels are related by exchange. 
Next we switch off the ZS channel. In other words, 
we study the running of the quasiparticle interaction
$F(\q')=m^\star/m \, f(\qp)$. This is shown in the right graph of
Fig.~\ref{toyflow}. Due to the factor $\Theta(q'-2\la)$ in the
beta function, the flow at small angles (small $q^\prime$) 
sets in only for small values of the cutoff $\la$.
\begin{figure}[t]
\begin{center}
\includegraphics[scale=0.27,clip=]{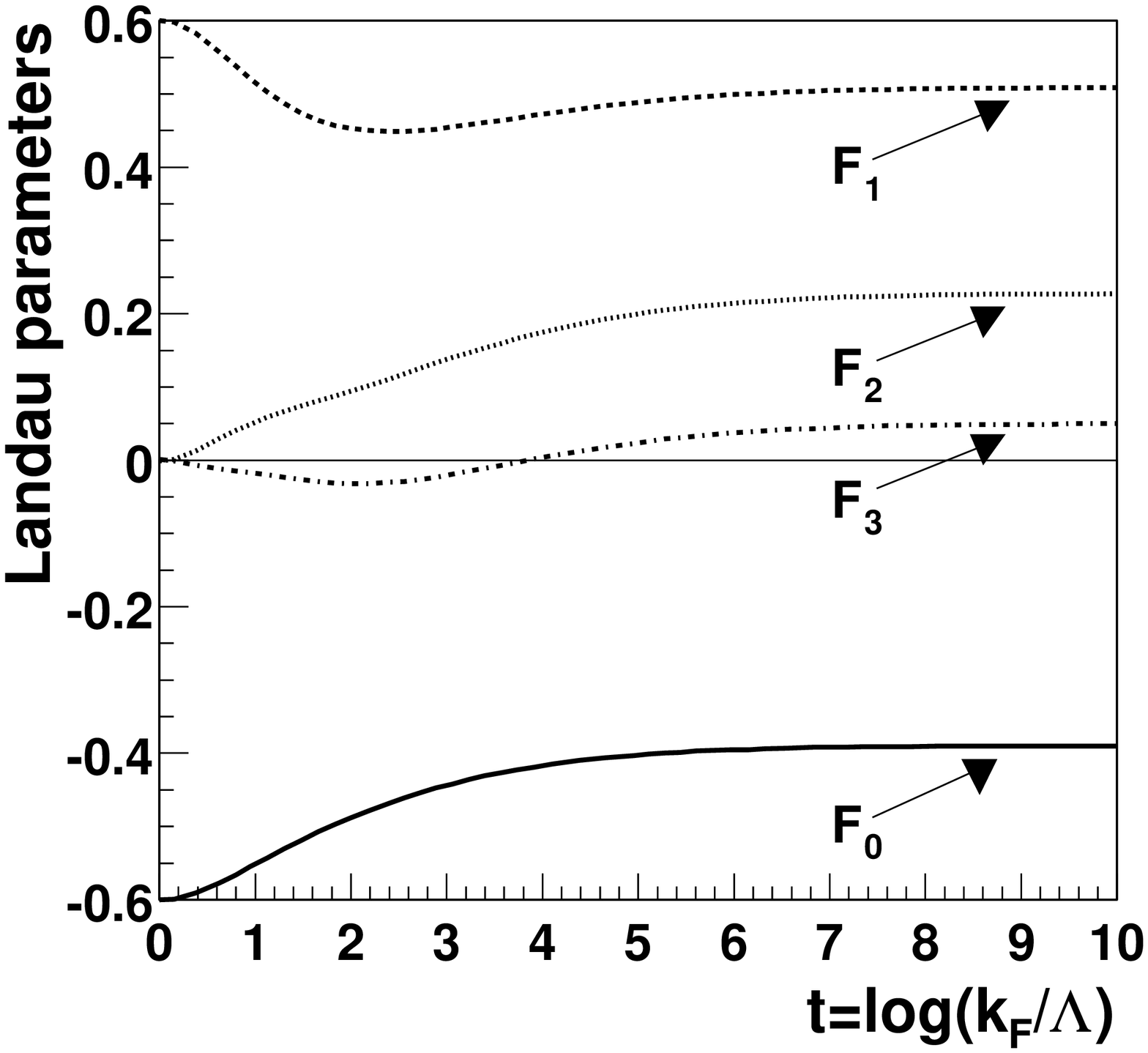}
\hspace*{0.5cm}
\includegraphics[scale=0.27,clip=]{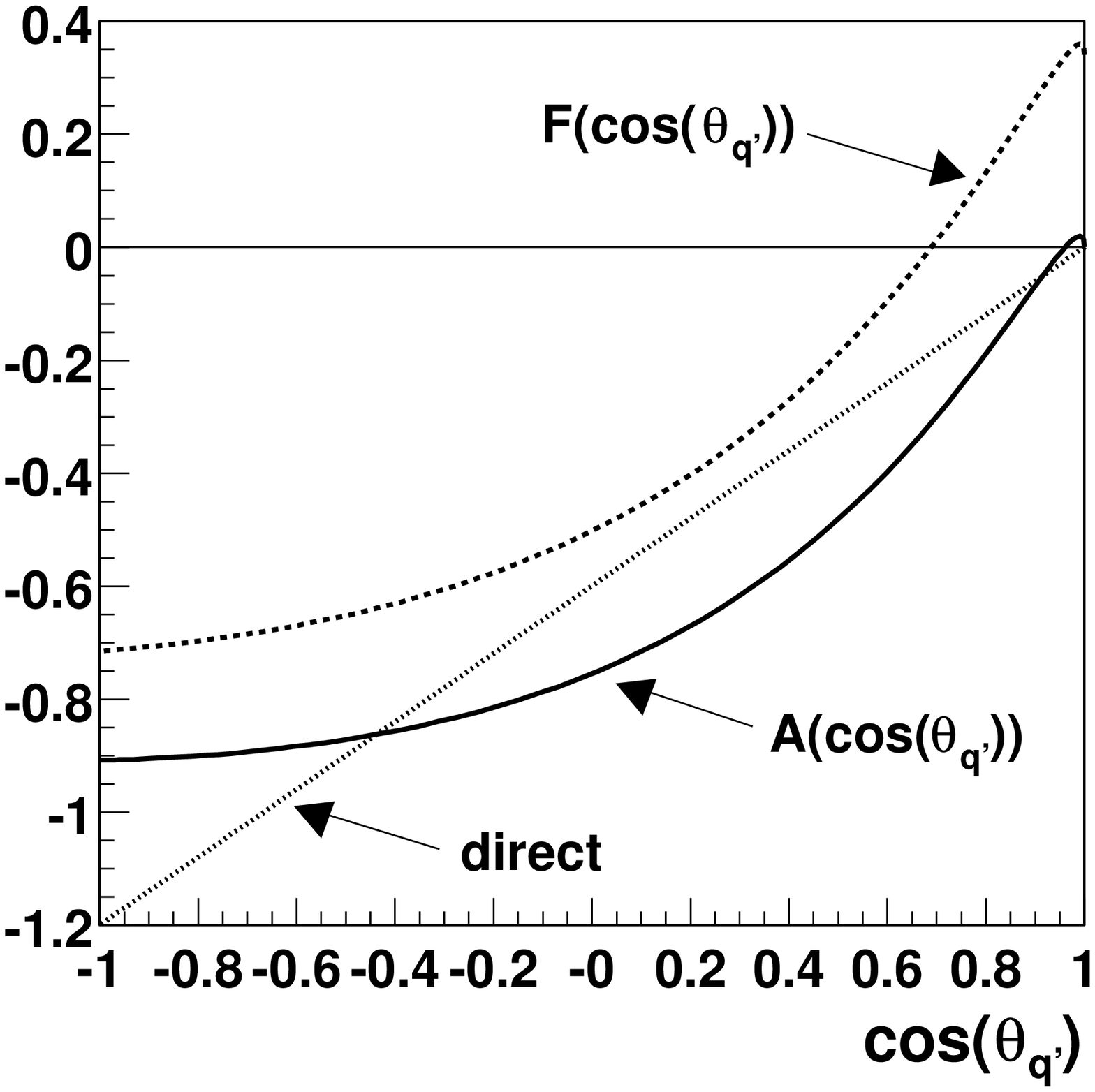}
\end{center}
\caption{The left figure shows the flow of the first four Landau
parameters. The right figure gives the solution for the quasiparticle 
interaction and the forward scattering amplitude and the 
initial ``direct'' term, Eq.~(\ref{direct}).}
\label{toysol}
\end{figure}

In Fig.~\ref{toysol}, we show the flow of the $l \leq 3$ Fermi
liquid parameters. The Fermi liquid parameters $F_l$ are obtained
by projecting the quasiparticle interaction on Legendre polynomials
$F(\cos\theta_\qp) = \sum_l F_l P_l(\cos\theta_\qp)$. The solution of
the RG flow naturally generates higher Fermi liquid parameters,
corresponding to higher powers of momentum transfer. Thus,
it is very convenient to solve the RG equations without projecting
on a set of coupling constants. The quasiparticle interaction and
the forward scattering amplitude obtained in the RG are shown in
Fig.~\ref{toysol}.

\begin{figure}[t] 
\begin{center}
\includegraphics[scale=0.27,clip=]{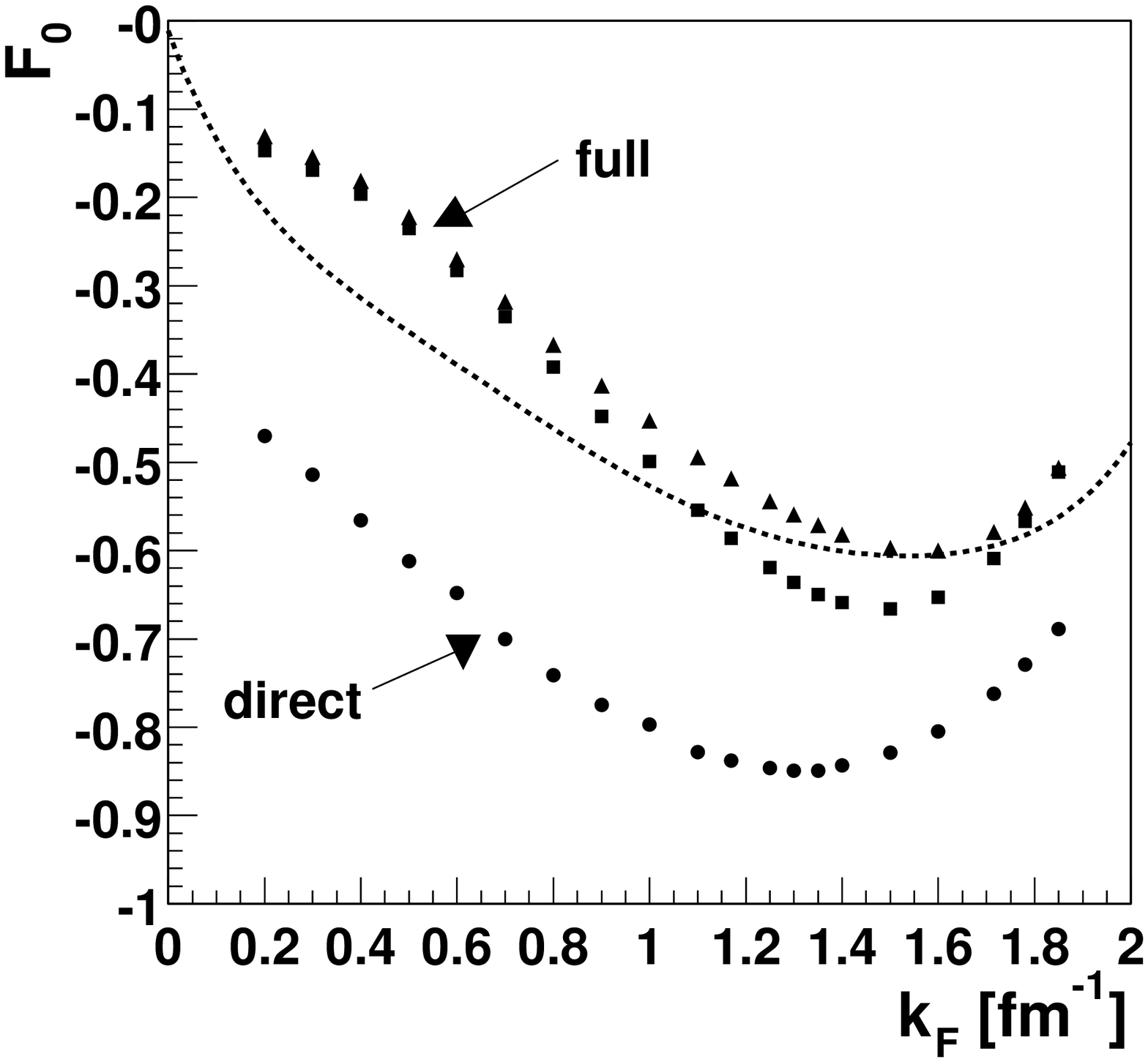}
\hspace*{0.5cm}
\includegraphics[scale=0.27,clip=]{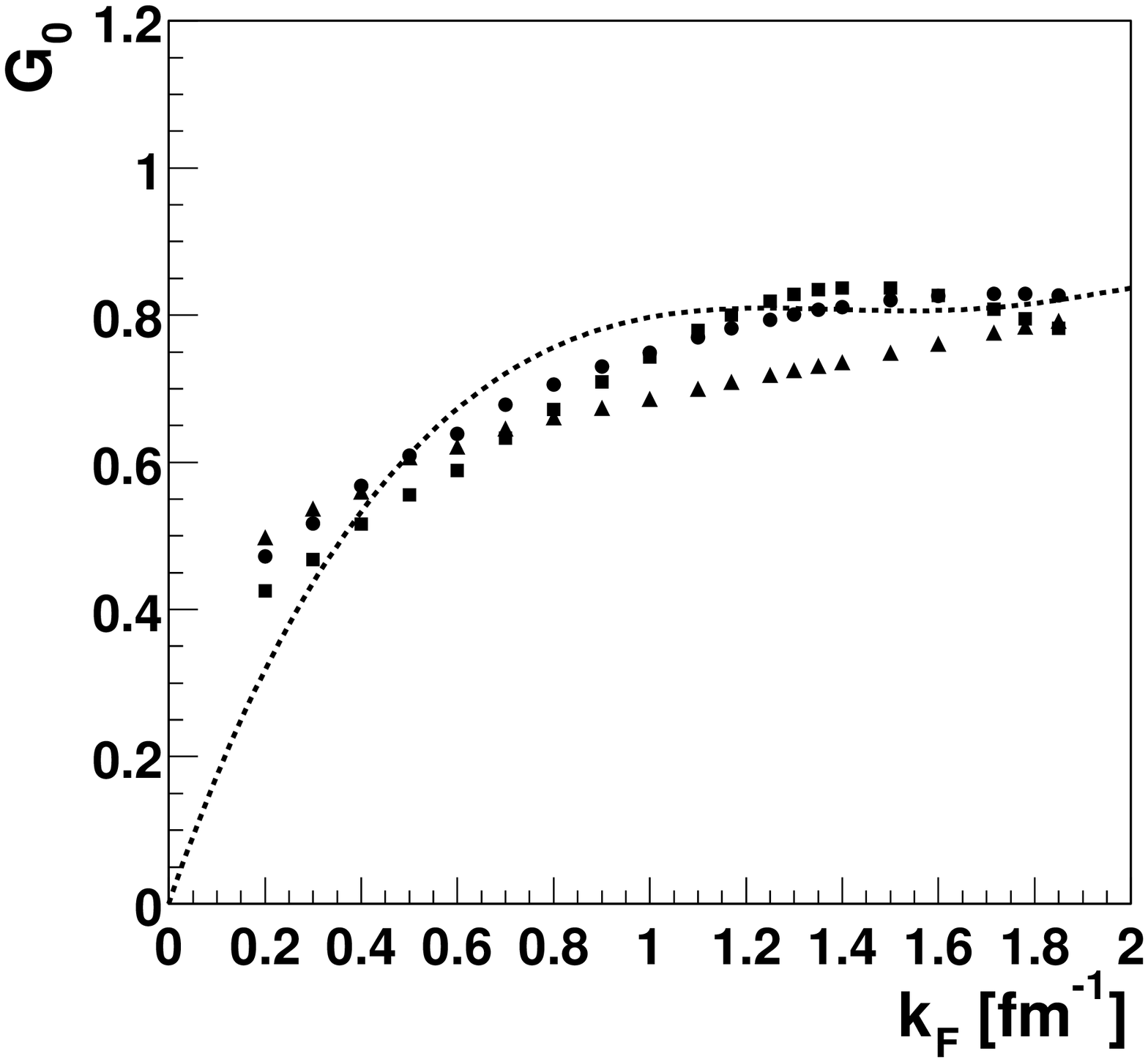}
\end{center}
\caption{The density dependence of the $l=0$ Fermi liquid
parameters of neutron matter $F_0$ (left figure) and $G_0$ 
(right figure), $\mathcal{F}(\cos\theta_\qp) = \sum_l ( F_l + G_l \,
\spin ) \, P_l(\cos\theta_\qp)$. Each graph contains the initial condition
of the flow (circles) and the full RG solution~\cite{RGnm} 
(squares for a density independent $z_{k_{\text{F}}}^2=0.9$ and 
triangles with an adaptive $z_{k_{\text{F}}}$ factor) in comparison 
with the study of Wambach {\em et al.}~\cite{WAP} (dashed line). For
details regarding the $z_{k_{\text{F}}}$ factor, see~\cite{RGnm}. Note
that for the toy model we have set $z_{k_{\text{F}}}=1$.}
\label{fgsol}
\end{figure}

The adaption of the method to a realistic system, like neutron matter is 
straightforward; we include spin and employ \vlk~as initial condition.
The details of the calculation are
given in~\cite{RGnm}. Here, we show only the $l=0$ Fermi liquid
parameters in comparison with the calculation of Wambach {\em
et al.}~\cite{WAP}. We note the importance of the large spin Fermi
liquid parameter $G_0 \approx 0.7-0.8$. It induces spin-density 
fluctuations which increase the incompressibility, related to
$1+F_0$, considerably, as can be seen in Fig.~\ref{fgsol}. Strikingly,
our results for the Fermi liquid parameters in this simple RG
calculation agree very well with Wambach {\em et al.}~\cite{WAP}.

In addition to Fermi liquid theory, the RG approach makes the full
scattering amplitude for finite $q$ accessible. The $q$ dependence
has been used to calculate the superfluid $^1$S$_0$ pairing gap
including polarization effects for neutron matter~\cite{RGnm}. We
conclude by remarking that the RG method is a very promising 
tool for studying a wide range of nuclear many-body problems.

\begin{ack}
This work was supported in part by the US-DOE grant No. DE-FG02-88ER40388.
\end{ack}

\end{document}